\documentclass[prb,floatfix,showkeys,twocolumn,10pt]{revtex4-1} 
\usepackage{amsmath,amsfonts,amssymb} 
\usepackage{graphicx} 
\usepackage{subfigure}
\usepackage[colorlinks,allcolors=blue]{hyperref}
\def\x{\end{document}}
\begin{document}

\title{Tunneling through a one-dimensional piecewise
constant potential barrier---{\small Illustrated with a non-uniform multi-barrier system and the genesis of the `Alias Effect'}}

\author{Siddhant Das}
\email{siddhantdas@yahoo.co.in\\ 
ORCID: 0000-0002-4576-9716}
\affiliation{Electronics and Communication Engineering, National Institute of Technology, Tiruchirappalli, TN, India.}
\date{\today}

\begin{abstract}
In this paper we look at transmission through one-dimensional
potential barriers that are piecewise constant. The Transfer Matrix
approach is adopted and a new formula is derived for multiplying long matrix
sequences that not only leads to an elegant representation of the wave function,
but also results in much faster computation than earlier methods. The proposed
method covers a broad spectrum of potentials of which multi-barrier systems are
special cases. The paradigm is exemplified with a finite lattice of non-uniform
rectangular barriers---non-uniformity being crucial, as the uniform case has been
solved exactly by Griffiths and Steinke. For the non-uniform multi-barrier
problem, the intervening wells strongly influence the transmission probability.
Surprisingly, we find that the wells act `individually', i.e. their influence is only a
function of their width and is independent of their exact `location' in a multi-barrier
system. This leads to a startling observation, which we have termed as the
`Alias Effect'. The exact solutions are supported with asymptotic formulas.
\keywords{
Piecewise Constant Potential, Resonant Multi-Barrier Tunneling, Transfer Matrix, Transmission
Coefficient, Alias Effect
}
\end{abstract}
\maketitle 

\section{Introduction} \label{sec1}

Tunneling of particles is a ubiquitous quantum phenomenon that gained lot of 
interest since its conception (by Hund\cite{sid1}) and is a subject of intense 
study even today. Many queer properties of matter can be understood based on 
the tunneling characteristics of charge carriers---for instance the 
emergence of band structure in solids. Moreover, tunneling in scanning 
tunneling microscopy, tunneling magnetic resistance, Josephson tunneling and 
many other physical phenomena rest directly on the transmission of particles 
through quantum barriers. It is rather difficult to obtain exact solutions 
of the Schr\"{o}dinger equation for arbitrary potentials which are of 
general interest. Thus one must consider the problems on a case by case 
basis. In this paper we look at a wide class of potential barriers that are 
piecewise constant. A piecewise constant potential barrier is 
discontinuous at one or more points. We will make the potential geometry 
more precise later. 

 The motivation for this study is twofold. First of all some potentials that 
arise in applications in condensed matter theory are special cases of this 
problem. For instance the analysis of semi-conductor super lattices by Tsu 
and Esaki\cite{sid2} was founded on a uniform multiple rectangular barrier model. A 
finite lattice of rectangular barriers is one of the simplest examples of a 
piecewise constant potential barrier. Although this problem was taken up by 
many researchers who presented analytic solutions for small number of 
barriers\cite{sid3,sid4,sid5}, Griffiths and Steinke\cite{sid6} have provided an exact solution to 
the problem for any number of barriers. However the method adopted in their 
paper does not extend to a lattice of \textit{non-uniform} barriers. As the presence of even 
mild asymmetry (in this problem) leads to very unusual quantum behavior\cite{sid11,sid9}, we have focused on asymmetry. We also show that the problem can be 
solved \textit{exactly} for any number of barriers. Secondly, the potentials of interest are 
smooth functions, for which analytic solutions can seldom be found. However 
the continuous potential can always be approximated to any level of accuracy 
as a sequence of flat steps. The resulting potential falls under the purview 
of the present class of problems, which can be solved exactly. The accuracy 
of these `step solutions' can be made arbitrarily good by choosing finer and 
finer partitions. Thus it is worthwhile to consider a piecewise constant potential barrier.

In the following section we formulate the problem precisely. By adopting the 
transfer matrix approach\cite{sid2} an explicit formula for multiplication of 
arbitrarily long matrix sequences is derived, thus obtaining the 
transmission characteristics of the potential \textit{exactly}. This forms the central part 
of the paper. In Section~\ref{sec3} the uniform multi-barrier is revisited and the 
role of asymmetry is demonstrated. The multi-barrier \textit{Alias Effect} is introduced and 
illustrated with examples in Section~\ref{sec3.1}. We conclude the analysis in Section~\ref{sec4}, outlining 
prospects of further study.

\section{Problem formulation} \label{sec2}

\begin{figure}[h!]
\centering
\includegraphics[scale=0.35]{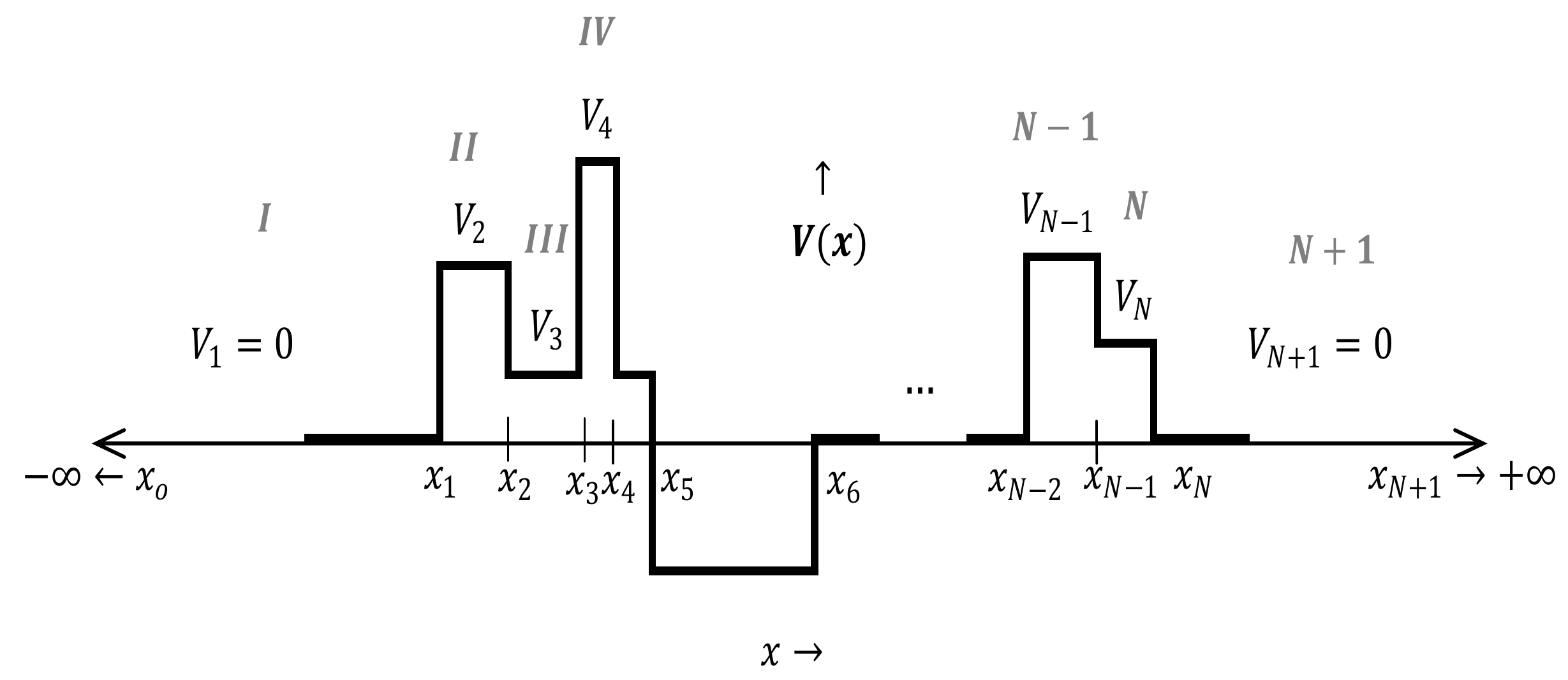} 
\caption{$V(x)$ for an arbitrary piecewise constant  potential barrier with $N$ jump discontinuities.}
\label{fig1}
\end{figure}

Figure~\ref{fig1} depicts a schematic piecewise constant potential barrier that 
requires the specification of two \textit{real valued} sequences for its definition. These are 
denoted by $\left\{x_{j} \right\}_{j=0}^{N+1} $ and $\left\{V_{j} 
\right\}_{j=1}^{N+1} $, where $N$ is the number of jump discontinuities of 
the potential barrier. $\left\{x_{j} \right\} $ must be an increasing 
sequence. In the following discussion $j$ runs from 1 to $N+1$ unless 
otherwise stated.

For consistency of notation we choose $x_{j} $ on the extended real line and 
require $x_{0} =-\infty $ and $x_{N+1} =+\infty $. For a localized barrier we let
 $V_{1} =V_{N+1} =0$. The potential $V(x) $ can be written as,
\begin{equation} \label{eq1}
 V(x)=V_{j},\quad x_{j-1} <x<x_{j}. 
\end{equation}

Figure~\ref{fig1} labels the regions of constant potential by Roman numerals. In these 
regions the Time Independent Schr\"{o}dinger Equation 
 has to be solved independently to yield the wave function 
$\psi (x)$, which is defined piecewise, 
\begin{equation}\label{eq2}
 \psi (x)=\psi_{j}(x),\quad x_{j-1} <x<x_{j}, 
\end{equation}
where the $\psi_{j} $s satisfy
\begin{equation} \label{eq3}
 \psi_{j}^{''} +\kappa_{j}^{2} \psi_{j} =0,\quad   \kappa 
_{j}^{2} \buildrel {\rm def} \over = \frac{2\mathfrak{M}\left(E-V_{j} \right)} {\hslash 
^{2}},
\end{equation}
$\mathfrak{M}$ is the mass of the particle with energy $E$. Equation~\eqref{eq3} admits general 
solutions of the form
\begin{equation} \label{eq4}
 \psi_{j} =A_{j} e^{i\kappa 
_{j} x} +B_{j} e^{-i\kappa_{j} x}. 
\end{equation}
$A_{j} $ and $B_{j} $ are the probability amplitudes for the forward and 
backward travelling wave components respectively. These amplitudes are 
collected in the ket $\vert \theta_j\rangle$.
\begin{equation}\label{eq5}
\vert \theta_j\rangle 
\buildrel {\rm def} \over = \binom{A_{j}}{B_{j}}.
\end{equation}
For the problem under consideration there exist $N+1$ tessellations of the 
real axis given by the intervals $x_{j-1} <x<x_{j},\ j=1, 
2,\ldots, N+1$. In each region $\psi $ is defined by equation~\eqref{eq4} up 
to two undetermined constants---a total of $2N+2$ constants. We are 
essentially looking at the transmission problem. Hence we assume that the 
particle is incident from left.\footnote{There is no loss of generality. One might as well choose right incidence. Since equation~\eqref{eq7} is a forward difference equation, it is 
amicable to left incidence calculations. As the transfer matrix is \textit{non singular}, one can 
readily invert it on to the other side, followed by a re-labeling of the 
index in equation~\eqref{eq7} (i.e. $j\to j-1$). This can then be 
used in an analogous fashion to handle the case of right incidence and 
$A_{1} $ must be set to zero (instead of $B_{N+1}$). Later, the Reflection and 
Transmission coefficients must be redefined as $\mathbb{T}=|\frac{B_{1}} {B_{N+1}}|^{2},\ 
\mathbb{R}=|\frac{A_{N+1}} {B_{N+1}} |^{2}$.}
 This input fixes one of the constants by requiring $B_{N+1} =0$ (since there 
is no reflector at $+\infty$). 
 
It now remains to determine the $2N+1$ constants. Any $2N$ of these can be 
uniquely expressed in terms of the other amplitude by necessitating that the 
wave function and its derivative be continuous at the discontinuities of 
$V(x)$. This also satisfies the equation of continuity. For illustrating the 
solution we choose $A_{N+1} $ to be the independent amplitude (in terms of 
which others are expressed) and this must be specified through an initial 
condition$ ({\Psi} \left(x,0 \right) = \psi 
(x)e^{-i\frac{E} {\hslash} t} \vert_{t = 0} $, for instance). 

 The required smoothness of the wave function is guaranteed by
\begin{equation}\label{eq6}
\left. 
{\begin{aligned}
\psi_{j} &=\psi_{j+1} \\
\frac{d\psi_{j}} {dx} &=\frac{d\psi_{j+1}} {dx} 
\end{aligned}} \right|_{x=x_{j}},\quad j=1,2,\ldots, N. 
\end{equation}
Equation~\eqref{eq6} translates into 
\begin{equation}\label{eq7}
 \vert \theta_j\rangle
=\mathbf{M}_{j} \vert \theta_{j+1}\rangle,
\end{equation}
where
\begin{widetext}
\begin{equation}\label{eq8}
 \mathbf{M}_{j} =\frac{1} {2\kappa_{j}} 
\begin{pmatrix}
\left(\kappa_{j} +\kappa_{j+1} \right)e^{i\left(\kappa_{j+1}-\kappa_{j}\right)x_{j}}&
\left(\kappa_{j}-\kappa_{j+1} \right)e^{-i\left(\kappa _{j} +\kappa_{j+1} \right)x_{j}} \\
\left(\kappa_{j}-\kappa_{j+1} \right)e^{i\left(\kappa_{j} +\kappa_{j+1} \right)x_{j}} & 
\left(\kappa_{j} +\kappa_{j+1} \right)e^{i\left(\kappa_{j}-\kappa_{j+1} \right)x_{j}} 
\end{pmatrix}. 
\end{equation}
\end{widetext}
$\mathbf{M}_{j} $s are known as the transfer matrices.\cite{sid5,sid7} Note that 
$\mathbf{M}_{j} $ is nonsingular for all $j$ (except when $E=V_{j+1})$ with a 
determinant ${\Delta}_{j} =2\kappa_{j+1} $. Moreover for $E=V_{j}$ 
(possibly for more than one $j$), ${\kappa}_{j} =0$ and 
$\mathbf{M}_{j} $ becomes indeterminate. We handle this case separately at 
the end of the section. The transfer matrices are rather special and are 
endowed with \textit{strong} algebraic properties which are ramifications of the continuity 
equation. We will return to this point later. By iterating equation~\eqref{eq7}, 
we express $\vert \theta_j\rangle
$ in terms of $ \vert \theta_{N+1}\rangle
=\binom{A_{N+1}}{0}$.
\begin{equation}\label{eq9}
 \vert \theta_j \rangle
=\mathbf{M}_{j} \mathbf{M}_{j+1} \cdots \mathbf{M}_{N} \vert 
\theta_{N+1}\rangle .
\end{equation}
Computation of the matrix product sequence in equation~\eqref{eq9} poses a 
formidable challenge, especially when $N$ is large.  If all the 
$\mathbf{M}_{j} $s were identical (which is the case for a uniform 
multi-barrier system), the product in equation~\eqref{eq9} reduces to a power, 
which can be calculated using well defined prescriptions. Griffiths and 
Steinke have exploited this advantage of the uniform multi-barrier problem. 
However (in general) all $\mathbf{M}_{j} $s would be different and a unified 
method must be outlined to efficiently handle long matrix product sequences. 
We overcome this problem by using an alternative representation for the 
$\mathbf{M}_{j} $s and then deriving a formula for the matrix product. 
 
 Any $2\times 2$ complex valued matrix can be uniquely expressed as a linear 
combination of the three Pauli matrices and the identity matrix, which 
collectively span $\mathbb{C}^{2\times 2} $. They are listed below.
\begin{equation}\label{eq10}
\begin{aligned}
\boldsymbol{\sigma}_{0} &=\begin{pmatrix}1 & 0\\0 & 1\\\end{pmatrix}, &
\boldsymbol{\sigma}_{1}  &=\begin{pmatrix}0 & 1\\1 & 0\\\end{pmatrix},\\
\boldsymbol{\sigma}_{2} &=\begin{pmatrix}0 &-i\\i & 0\\\end{pmatrix}, &
\boldsymbol{\sigma}_{3} &=\begin{pmatrix}1 & 0\\0 &-1\\\end{pmatrix}. 
\end{aligned}
\end{equation}
The collection $\left\{\boldsymbol{\sigma}_{p} \right\} $ is the Pauli basis. 
Now,
\begin{equation}\label{eq11}
 \mathbf{M}_{j} =\sum\limits_p {c_{j}^{p} \boldsymbol{\sigma}_{p}},\quad 
 c_{j}^{p} =\frac{1} {2} {\rm trace} \left(\mathbf{M}_{j} \boldsymbol{\sigma 
}_{p} \right).
\end{equation}
In this form $\mathbf{M}_{j} $ is identified as a \textit{Pauli Vector}. The subscript $j$ in the 
scalar $c_{j}^{p}$ denotes the order of the transfer matrix while the 
superscript is identified with the index of the basis element it is 
multiplied with. In all the summations that follow the index runs over $0, 
1, 2, 3$ unless otherwise stated. It can be shown that the 
product of two transfer matrices,
\begin{align}
 \mathbf{M}_{j} \mathbf{M}_{k} &=\left(\sum\limits_p {c_{j}^{p} \boldsymbol{\sigma 
}_{p}} \right)\left(\sum\limits_q {c_{k}^{q} \boldsymbol{\sigma}_{q}} 
\right)\nonumber\\[1ex]
&=\sum\limits_p {\boldsymbol{\sigma}_{p} \sum\limits_q 
{c_{j}^{q} c_{k}^{\phi \left(p,q \right)} (i)^{\varepsilon_{p 
q \phi \left(p,q \right)}}}} \label{eq12}
\end{align}
where,
\begin{align}
 \phi \left(a,b \right)&= \left(a+{b\left(-1 \right)}^{a+b-1} 
\right)\mod{4}\label{eq13}\\
 \varepsilon_{a b c} &=\frac{1} {2} \left(a-b \right)\left(b-c 
\right)\left(c-a \right).\label{eq14}
\end{align}

Equation~\eqref{eq12} results from expanding the bracketed pair and injecting the 
product identities of the Pauli matrices.\cite{sid8} $\varepsilon_{a b c} $ is 
the Levi-Civita Symbol (or permutation symbol) which along with $\phi \left(
a,b \right)$ preserves the non-commutativity of matrix multiplication. 
Equation~\eqref{eq12} expresses $\mathbf{M}_{j} \mathbf{M}_{k} $ in the form of 
equation~\eqref{eq11}, which is a distinctive advantage 
since \textit{matrix products get expressed as linear combinations of simple matrices}. 

We use induction to obtain a Pauli Vector representation for the matrix 
product sequence appearing in equation~\eqref{eq9}. In the following discussion 
the summation indices are augmented with an additional subscript for the 
sake of clarity. The recipe holds good for multiplying arbitrary $2\times 2$ 
matrices (not necessarily transfer matrices). We illustrate the inductive 
construction by multiplying $m$ matrices $\left\{\mathbf{T}_{j} \in 
\mathbb{C}^{2\times 2} \right\}_{j=1}^{m} $ beginning with $\mathbf{T}_{1}$,
\begin{widetext}
\begin{align*}
\mathbf{T}_{1} &=\sum_{p_{1}} \boldsymbol{\sigma}_{p_{1}} c_{1}^{p_{1}}\\
\mathbf{T}_{1} \mathbf{T}_{2} &=\sum\limits_{p_{2}} \boldsymbol{\sigma}_{p_{2}} 
c_{12}^{p_{2}}=\sum\limits_{p_{2}} \boldsymbol{\sigma}_{p_{2}} \sum\limits_{q_{1}} 
{c_{1}^{q_{1}} c_{2}^{\phi \left(p_{2},q_{1} \right)}} (i)^{\varepsilon_{p_{2} q_{1} \phi \left(p_{2},q_{1} \right)}} \\
\left(\mathbf{T}_{1} \mathbf{T}_{2} 
\right)\mathbf{T}_{3} &=\sum\limits_{p_{3}} \boldsymbol{\sigma}_{p_{3}} 
c_{123}^{p_{3}} =\sum\limits_{p_{3}} \boldsymbol{\sigma}_{p_{3}} \sum\limits_{q_{2}} 
{c_{12}^{q_{2}} c_{3}^{\phi \left(p_{3},q_{2} \right)}} (i)^{\varepsilon_{p_{3} q_{2} \phi \left(p_{3},q_{2} \right)}} \\
&=\sum\limits_{p_{3}} \boldsymbol{\sigma}_{p_{3}} \sum\limits_{q_{2}} 
{\sum\limits_{q_{1}} {c_{1}^{q_{1}} c_{2}^{\phi \left(q_{2},q_{1} \right)}} 
c_{3}^{\phi \left(p_{3},q_{2} \right)}} (i)^{\varepsilon 
_{q_{2} q_{1} \phi \left(q_{2},q_{1} \right)} +\varepsilon_{p_{3} 
q_{2} \phi \left(p_{3},q_{2} \right)}} \\
\left(\mathbf{T}_{1} \mathbf{T}_{2} \mathbf{T}_{3} 
\right)\mathbf{T}_{4} &=\sum\limits_{p_{4}} \boldsymbol{\sigma}_{p_{4}} 
c_{1234}^{p_{4}} =\sum\limits_{p_{4}} \boldsymbol{\sigma}_{p_{4}} 
\sum\limits_{q_{3}} {c_{123}^{q_{3}} c_{4}^{\phi \left(p_{4},q_{3} \right)}} 
(i)^{\varepsilon_{p_{4} q_{3} \phi \left(p_{4},q_{3} 
\right)}} \\
 &=\sum\limits_{p_{4}} \boldsymbol{\sigma}_{p_{4}} \sum\limits_{q_{3}} \sum\limits_{q_{2}} 
\sum\limits_{q_{1}} c_{1}^{q_{1}} c_{2}^{\phi \left(q_{2},q_{1} \right)} c_{3}^{\phi \left(
q_{3},q_{2} \right)} c_{4}^{\phi \left(p_{4},q_{3} \right)} (i)^{\varepsilon_{q_{2} q_{1} \phi \left(q_{2},q_{1} \right)} +\varepsilon_{q_{3} q_{2} \phi \left(q_{3},q_{2} 
\right)} +\varepsilon_{p_{4} q_{3} \phi \left(p_{4},q_{3} \right)}} 
\end{align*}
setting $q_{0} \buildrel {\rm def} \over = 0 $ and noting that $q_{1} =\phi 
\left(q_{1},0 \right)=\phi \left(q_{1},q_{0} \right)$,  the general formula 
can be written as 
\begin{align}\label{eq15}
&\mathbf{T}_{1} \mathbf{T}_{2} \cdots 
\mathbf{T}_{m-1} \mathbf{T}_{m}\nonumber\\
&=
\sum\limits_{p_{m}} \boldsymbol{\sigma}_{p_{m}} 
\sum\limits_{q_{m-1}} 
\sum\limits_{q_{m-2}} 
\cdots 
\sum\limits_{q_{3}} 
\sum\limits_{q_{2}} 
\sum\limits_{q_{1}} 
\prod\limits_{j=1}^{m-1}
\left(c_{j}^{\phi \left(q_{j},q_{j-1} 
\right)} (i)^{\varepsilon_{q_{j} q_{j-1} \phi \left(
q_{j},q_{j-1} \right)}} \right) c_{m}^{\phi \left(p_{m},q_{m-1} 
\right)} (i)^{\varepsilon_{p_{m} q_{m-1} \phi \left(
p_{m},q_{m-1} \right)}}.\nonumber\\
\end{align}
\end{widetext}
\noindent
It must be stated that the representation depicted above is not unique. For 
instance, in calculating the product of four matrices one can use 
\textit{associativity} to multiply two of these in two pairs and use equation~\eqref{eq12} to compose 
the resulting pair---in which case a different form would result involving 
higher compositions of $\varepsilon_{abc} $ and $\phi (a,b)$. Naturally, these 
are equivalent representations. However we chose the form given in equation~\eqref{eq15} for its compact representability and ease of computation. Note that 
in equation~\eqref{eq15} except for the outer most summation, the inner multiple 
summations are \textit{scalars}. Although the computational power of equation~\eqref{eq15} is not 
readily apparent, it can be used to solve the problem  for any $N$ in 
a reasonable time. It will also turn out to be a useful manual aid for 
obtaining closed form solutions for small barrier numbers which otherwise 
require lot of effort. 

We return to the required product sequence of equation~\eqref{eq9} by mapping 
$\mathbf{T}_{1} \to \mathbf{M}_{j} $, $\mathbf{T}_{2} \to \mathbf{M}_{j+1} $, 
$\mathbf{T}_{3} \to \mathbf{M}_{j+2}$, \ldots, $\mathbf{T}_{m} \to \mathbf{M}_{N} $ 
in equation~\eqref{eq15}. To avoid the long formula we denote this product as
\begin{equation}\label{eq16}
\mathbf{M}_{j} \mathbf{M}_{j+1} \mathbf{M}_{j+2} \ldots 
\mathbf{M}_{N-1} \mathbf{M}_{N} = \sum\limits_p {\mu_{j}^{p} \boldsymbol{\sigma 
}_{p}},
\end{equation}
where the $\mu_{j}^{p}$s can be readily obtained using the above 
prescription. The set of transfer matrices $\left\{\mathbf{M}_{j} 
\right\}_{j=1}^{N} $ is independent of initial conditions, and is completely 
specified by the barrier parameters and particle energy, through equation~\eqref{eq8}. 
Thus we have uniquely expressed every $\vert 
\theta_j \rangle
$ in terms of $\vert \theta_{N+1}\rangle
=A_{N+1} \vert +\rangle$, where $\vert + \rangle=\binom{1}{0}$ and  $\vert - \rangle=\binom{0}{1}$. With equation~\eqref{eq9}, \eqref{eq16} and an initial 
condition, $\psi (x)$ is pinned down uniquely. 
And equation~\eqref{eq4} can be rewritten as 
\begin{align}
 \psi_{j} (x)&=\langle + | \theta_j \rangle e^{i\kappa_{j} x} +\langle -| 
\theta_j \rangle e^{-i\kappa_{j} x} \nonumber\\
&=\langle + \vert \left(\prod\limits_{l=j}^N \mathbf{M}_{l} \right)\vert 
\theta_{N+1} \rangle e^{i\kappa_{j} x}\nonumber\\&\qquad + \langle - 
\vert \left(\prod\limits_{l=j}^N \mathbf{M}_{l} \right)\vert 
\theta_{N+1} \rangle	e^{-i\kappa_{j} x} \nonumber\\
&=\langle + \vert \left(\sum\limits_p {\mu_{j}^{p} \boldsymbol{\sigma}_{p}} \right)\vert 
\theta_{N+1} \rangle e^{i\kappa_{j} x}\nonumber\\&\qquad +\langle - 
\vert \left(\sum\limits_p {\mu_{j}^{p} \boldsymbol{\sigma}_{p}} \right)\vert 
\theta_{N+1} \rangle e^{-i\kappa_{j} x} \nonumber\\
& =A_{N+1} \sum\limits_p {\mu_{j}^{p} \left(
 \langle + \vert  \boldsymbol{\sigma}_{p} \vert + \rangle
e^{i\kappa_{j} x} +\langle - \vert  \boldsymbol{\sigma}_{p} 
\vert + \rangle
e^{-i\kappa_{j} x} \right)} \nonumber\\
& =\left(\mu_{j}^{0} +\mu_{j}^{3} 
\right)A_{N+1} e^{i\kappa_{j} x} +\left(\mu_{j}^{1} +i\mu_{j}^{2} 
\right)A_{N+1} e^{-i\kappa_{j} x} . \label{eq17}
\end{align}
Injecting equation~\eqref{eq17} in equation~\eqref{eq2} gives
\begin{align}\label{eq18}
 \psi (x)&=\left(\mu_{j}^{0} +\mu_{j}^{3} 
\right)A_{N+1} e^{i\kappa_{j} x} +\left(\mu_{j}^{1} +i\mu_{j}^{2} 
\right)A_{N+1} e^{-i\kappa_{j} x},
\nonumber\\&\qquad x_{j-1} <x<x_{j},\  
 j=1,2,\ldots, N+1.
\end{align}
At this point we address a problem that arises when $\kappa_{j} =0$ in 
equation~\eqref{eq8}. Note that this occurs when $E=V_{j} $ and equation~\eqref{eq3} 
leads to solutions of the form $\psi_{j} =A_{j} x+B_{j} $ (not exponentials). 
Using these in equation~\eqref{eq6} gives the correct transfer matrices and the 
subsequent procedure is same as before. The transmission and reflection 
coefficients can be readily calculated from equation~\eqref{eq18}. These are 
defined as\cite{sid7} 
\begin{equation}\label{eq19}
 \mathbb{T}=\left|  \frac{A_{N+1}} {A_{1}} \right|^{2}, \quad 
 \mathbb{R}=\left|  \frac{B_{1}} {A_{1}} \right|^{2} .
\end{equation}
The choice of letting $A_{N+1} $ be the independent undetermined constant was 
made considering the form of equation~\eqref{eq19}. Also note that these 
coefficients are independent of the  initial condition ($A_{N+1})$. 
Thus the complete set of transfer matrices uniquely determines 
the transmissibility of the potential barrier. Quite independent of the 
barrier geometry, an important identity follows from the equation of 
continuity for the probability current density: $\mathbb{T}+\mathbb{R}=1$.\footnote{This result can be derived explicitly from the continuity equation. A 
particularly neat (and equivalent) way of deriving the same is from the 
conservation of the average momentum $<p>$ associated with the wave function $\psi$.\cite{sid9} Note: The momentum associated with the wave function(s) $A_{\pm} e^{\pm i\kappa x} $ is  $\pm 
\hslash \kappa $. $A_{\pm} $ being the probability amplitude of finding the 
particle with $p=\pm \hslash \kappa $. Hence the average momentum associated 
with $\psi_{j}$ (equation~\eqref{eq2})  is 
$<\!p\!>
= \left(+\hslash \kappa_{j} \right)A_{j} A_{j}^{\ast} + 
\left(-\hslash \kappa_{j} \right)B_{j} B_{j}^{\ast} = \hslash \kappa 
_{j} \left(| A_{j} |^{2}-| B_{j} |^{2} \right) = 
\mathrm{const}\cdot  \kappa_{1} =\kappa_{N+1}$. 
Thus $| A_{1} 
|^{2}-| B_{1} |^{2} = | A_{N+1} |^{2}-| 
B_{N+1} |^{2} $. Since $B_{N+1} =0$, $\frac{| A_{N+1} 
|^{2}} {| A_{1} |^{2}} +\frac{| B_{1} 
|^{2}} {| A_{1} |^{2}} =1$ or $\mathbb{T}+\mathbb{R}=1$.}

The 
satisfaction of this identity constrains the individual transfer matrices, 
inducing \textit{strong} algebraic properties amidst its elements. Conversely, the matrices 
that satisfy these properties can only be transfer matrices for some 
tunneling problem. Griffiths and Steinke have derived some of these 
properties in their paper.\cite{sid6} Merzbacher also discusses these properties in 
his book.\cite{sid7} These authors discuss the properties possessed by individual 
transfer matrices. However, one important fact deserves appreciation. 
The `local' algebraic constraints amidst the elements of the individual 
transfer matrices ($\mathbf{M}_{j} $s) manifest in a \textit{similar} `global' identity for the 
transfer matrix product $\prod \mathbf{M}_{j}$ (of equation~\eqref{eq9}). This 
fact has not been recognized in the papers devoted to this problem. Moreover 
the existence of these properties for the product matrix is \textit{independent of the order of multiplication} of the transfer 
matrices. i.e. the induction of global product properties for $\prod 
\mathbf{M}_{j} $ from that of the individual transfer matrices 
$\mathbf{M}_{j} $s, overlooks the \textit{non-commutativity} of matrix multiplication! Note that the 
product matrix can be a very complicated object, depending on how many 
matrices are being multiplied. Irrespective of that, the global identities 
hold true and can be rigorously proven. These relationships are rather 
profound and we reserve a thorough discussion of the same for a future 
paper. 

We conclude this section by formulating $\mathbb{T}$ and $\mathbb{R}$ in terms of the 
Pauli coefficients of the transfer matrix product in equation~\eqref{eq20}.
\begin{equation}\label{eq20}
 \mathbb{T}=\frac{1} {\vert  \mu_{1}^{0} +\mu_{1}^{3} \vert^{2}},\quad 
 \mathbb{R}=\left|\frac{\mu_{1}^{1} +i\mu_{1}^{2}} {\mu_{1}^{0} +\mu 
_{1}^{3}} \right|^{2}.
\end{equation}
The fact that only \textit{two} coefficients---$\mu_{1}^{0} $ and $\mu_{1}^{3} $ show up 
in the expression for $\mathbb{T}$, in terms of which $\mathbb{R}$ can be readily expressed 
($\mathbb{R}=1-\mathbb{T})$ implies that the $\mu_{1}^{p} $s are not independent of each 
other. This again is an offshoot of the special transfer matrix properties 
mentioned above. At any rate, this signals a computational advantage---i.e. only $ \mu_{1}^{0} $ and $ \mu_{1}^{3} $ have to 
be found for computing $\mathbb{T}$.

\section{Discussion} \label{sec3}

We revisit the problem of tunneling through a finite lattice of \textit{uniform} rectangular 
barriers. This is a special case of a piecewise constant potential barrier. 
We are mainly interested to look at the transmission coefficient $\mathbb{T}$, in the 
situations when the barrier ceases to be uniform i.e. an asymmetric 
multi-barrier. The uniform barrier thus serves as a basis for comparison. In 
what follows, we prefer to work in units, where $\frac{2\mathfrak{M}} {\hslash^{2}} =1$. 
 
\begin{figure}[h!]
\centering
\subfigure{(a)}{\includegraphics[scale=0.35]{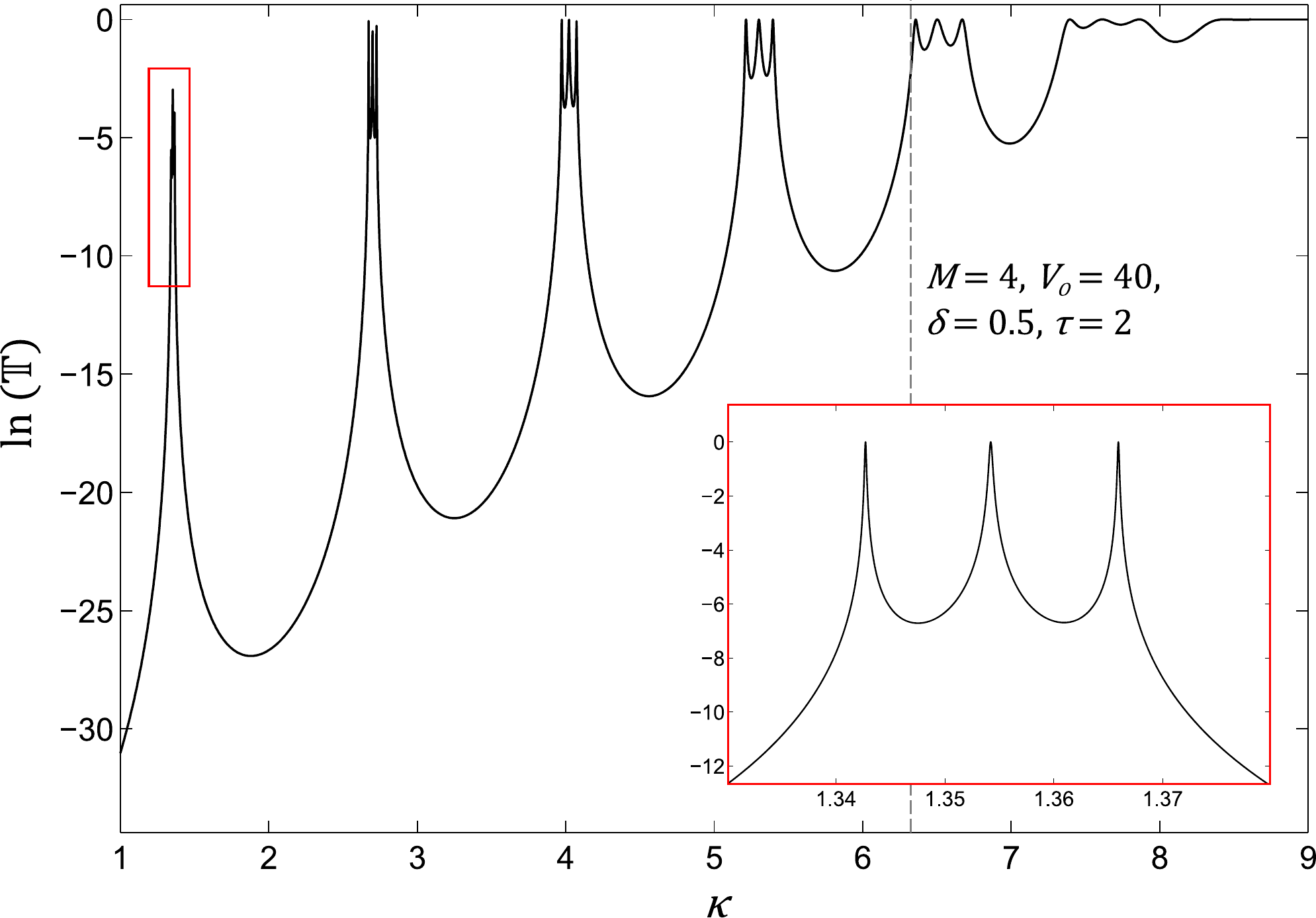}}
\subfigure{(b)}{\includegraphics[scale=0.35]{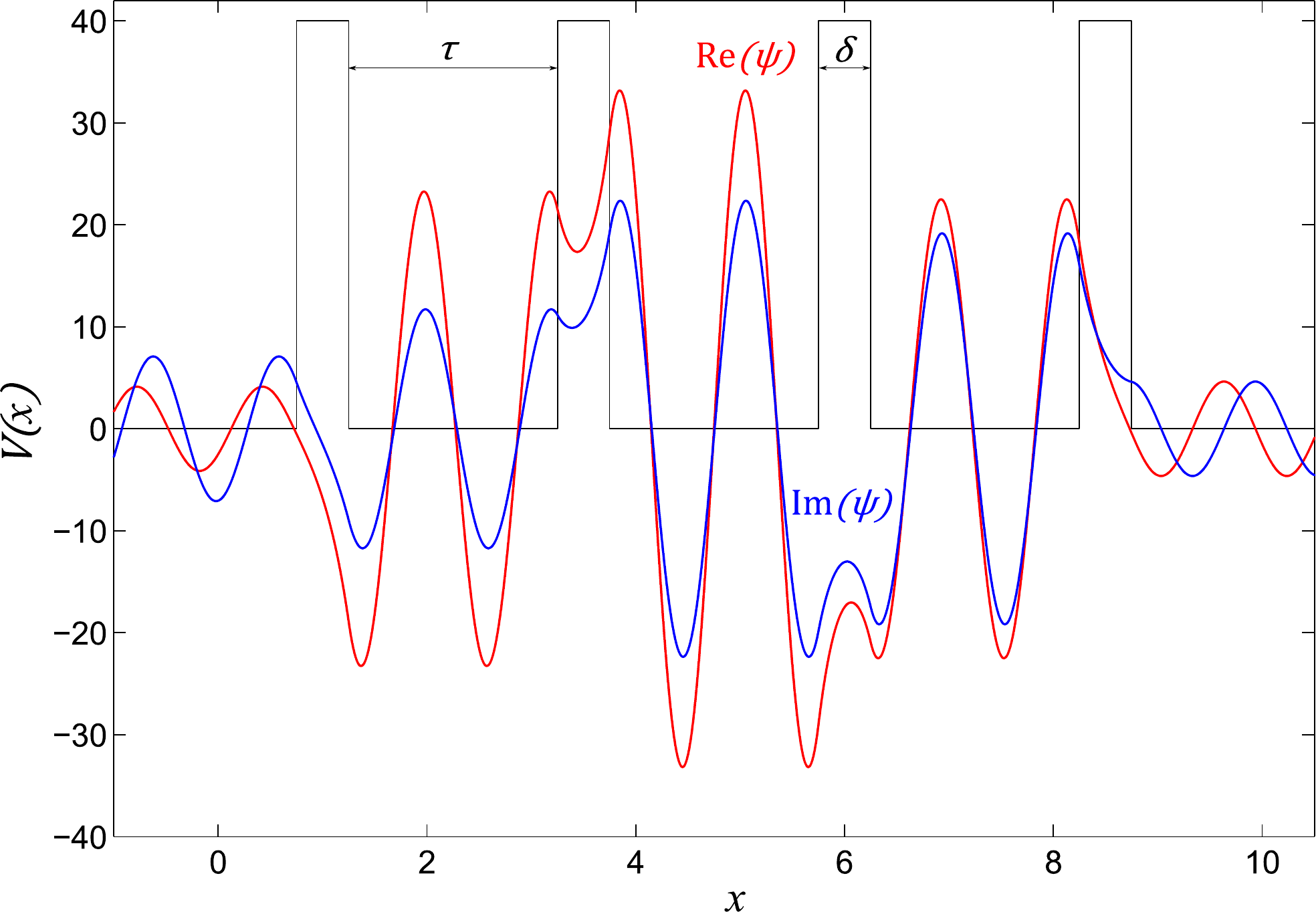}}
\caption{(a) $\ln(\mathbb{T})$ vs. $\kappa$ for a uniform 4BP with specifications: $V_0=40,\ \delta =0.5, \tau=2$. The broken vertical line denotes $\kappa =\sqrt{V_0} \sim  6.32$. (b) $V(x)$ for the same barrier, juxtaposed with the real and imaginary parts of $\psi(x)$ for a typical particle energy of 27.217 (in the chosen units). Note that this energy corresponds to the resonant peak at $\kappa=5.217$ of Fig.~\ref{fig2}(a). Further, $A_{N+1}=A_9$ is taken as 1, and the wave function is suitably scaled (by $\frac{V_0}{\max{|\psi(x)|}})$ to pose it along with the barrier.}
\label{fig2}
\end{figure}

 A collection of $m$ rectangular barriers constitutes a \textit{M Barrier Problem} or MBP. The 
potential $V(x)$ for a uniform MBP is specified with the sequences 
$\{x_{j}\}$ and $\{V_{j}\} $,
\begin{equation}\label{eq21}
\begin{aligned}
\{x_{j}\} &=
\begin{cases}
\frac{\left(j-1 \right)} {2} \left(\delta +\tau \right)+{\Theta 
}& j=1,3,5,\ldots, 2m-1\\
\frac{j} {2} \delta +\left(\frac{j} {2}-1 \right)\tau +{\Theta} 
 &j=2,4,6,\ldots, 2m\\
\end{cases}\\
\left\{V_{j} \right\} &=\begin{cases}
0 & j=1,3,5,\ldots, 2m-1\\ 
V_{0} & j=2,4,6,\ldots, 2m
\end{cases}
\end{aligned} 
\end{equation}
where $\delta $ is the barrier width, $\tau $ is the well width and $ 
V_{0} $ is the barrier height. $\Theta $ denotes the starting point of the 
barrier train, which could be conveniently shifted to zero as the physical properties 
of the problem are invariant to translation. The jump discontinuity number 
$N$ equals $2m$ for a MBP. Note that the barrier length $ L=m\delta 
+(m-1)\tau $. With $N=2m$ these sequences satisfy the requirements 
necessitated in Section II. As an example consider the case of $m=4,{
V}_{0} =40, \delta =0.5, \tau =2$. We plot $ \ln (\mathbb{T})$ vs. $\kappa (=\sqrt E $ in the prescribed units) for this barrier in Fig.~\ref{fig2}(a).\footnote{All the plots in this paper have been generated from programs 
written in MATLAB$^{\circledR}$ Version---7.12.0.635 (R2011a).}
$V(x)$ for this case is graphed in Fig.~\ref{fig2}(b) 
along with the real and imaginary parts of $\psi (x) $ for a typical energy 
of 27.217 (in the chosen units). For plotting the wave function we have 
chosen $A_{9}$ (i.e. $A_{N+1} =1$). $\psi $ has been scaled by a factor 
of $\frac{V_{0}} {\max (\vert  \psi (x) \vert )}$ to pose it along with the barrier. This also takes care 
of dimensions.

Note the presence of $3\ (m-1$, in general)\cite{sid3,sid5,sid6,sid9,sid10} resonant 
peaks in Fig.~\ref{fig2}(a), which are almost superposed on each other for $\kappa 
\ll \sqrt V_{0} $ and gradually resolve with increasing $\kappa $. The 
resonant peaks are grouped into distinct bands. Even for small $m$, the 
emergence of band structure is readily apparent, though this is more 
pronounced in the case of strictly periodic potentials.\cite{sid12} The resonances 
\textit{nearly} correspond to the bound states of the infinite square well of width $\tau 
$. These are given by
\begin{equation}\label{eq22}
 \kappa \tau =n\pi, \quad  n=1,2,3,\ldots 
\end{equation}
Since the barriers are of finite height, only at low energies ($E\ll 
V_{0})$, equation~\eqref{eq22} is a faithful estimator of the actual resonant 
energies. This correspondence starts to deviate as we move towards the 
barrier top, i.e. increase $n$. Rather unexpectedly, It will turn out that 
any MBP can be completely portrayed on the basis of equation~\eqref{eq22}. We 
compare the estimates obtained from equation~\eqref{eq22} with the exact 
solution (equation~\eqref{eq20}) in the following discussion.  
 An interesting result that can be deduced from equation~\eqref{eq22} is the 
maximum number of resonant bands (to be called as $\beta)$ that occur for 
$\kappa <\sqrt V_{0} $. This is obtained from the condition $\frac{n\pi 
} {\tau} \le \sqrt V_{0} $, which gives 
\begin{equation}\label{eq23}
 \beta =\left[\frac{\tau \sqrt V_{0}} {\pi} \right], 
\end{equation}
[ ] is the greatest integer (floor) function. For the 4BP of Fig.~\ref{fig2}(a), 
$\frac{\sqrt V_{0}} {\pi} \sim 2$ and the well width $\tau =2$. Thus $\beta 
=4$, (i.e. 4 resonant bands) which is true. We use the terms bands and 
peaks interchangeably at times, especially when the bands are very narrow. 
But it must be understood that the number of bands is $\beta $ and each band 
contains $m-1$ resonant peaks.\footnote{Only the uniform MBP admits well defined bands (A band being a local 
group of $m-1$ resonant peaks). R. Gilmore in his book\cite{sid9} terms these 
bands as $N $\textit{tuplets}, $N=m-1$. The $\beta $ of equation~\eqref{eq23} essentially gives 
the total \textit{number of bands} of a uniform MBP (below $V_{0})$. For an 
asymmetric MBP, the band structure is completely lost and is replaced by a 
stray collection of resonant peaks. $\alpha $ (defined in equation~\eqref{eq24}) 
gives the \textit{total number of resonant peaks} for an asymmetric MBP. Thus for a 
uniform MBP, $\alpha =(m-1)\beta $.} 
 
Arguably, the $m-1$ wells of a MBP, \textit{each} contribute a resonant peak (thus $m-1$ 
peaks) to each band below $V_{0} $. When the wells have the same width, the 
$m-1$ states in each band are degenerate (when the barrier height is 
infinity). For finite barrier height these levels couple, which leads to 
splitting of these $m-1$ levels. However the coupling is smallest for the 
lowest energy band and largest for the highest energy band (below $V_{0})$. 
Thus, the degeneracy is lifted only at higher energies.

\begin{figure}[h!]
\centering
\subfigure{(a)}{\includegraphics[scale=0.3]{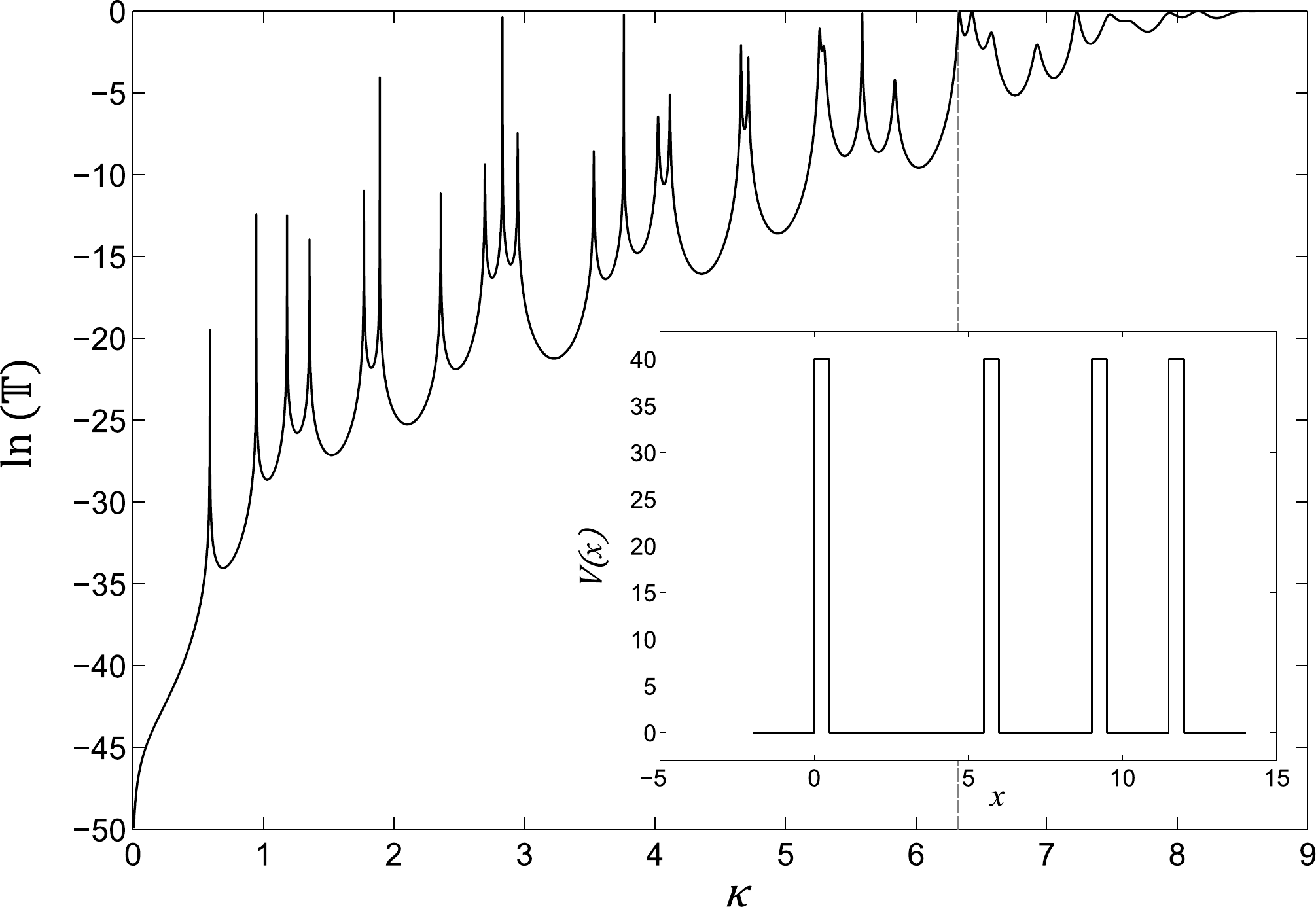}}
\subfigure{(b)}{\includegraphics[scale=0.25]{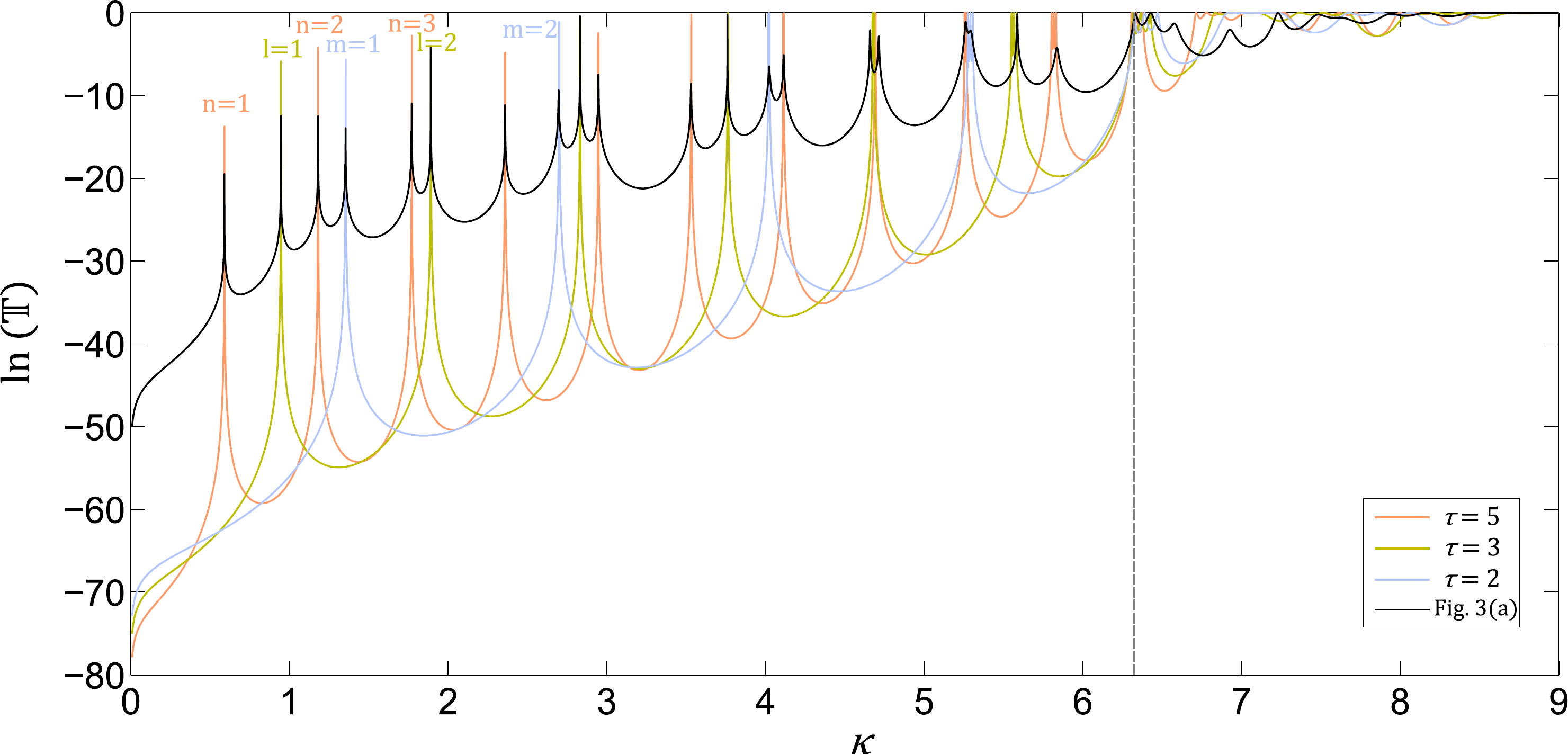}}
\caption{(a) $\ln(\mathbb{T})$ vs. $\kappa$ for an asymmetric 4BP with $V_0=40,\ \delta =0.5,\ \tau_1=5,\ \tau_2=3,\ \tau_3=2$. $V(x)$ is shown as inset. (b) Black curve is same as that of Fig.~\ref{fig3}(a). Red, green and blue curves plot  $\ln(\mathbb{T})$   vs. $\kappa$ for \textit{uniform} 4BPs with specifications: $V_0=40,\ \delta=1$ (same for all three curves) but different well widths (listed in the legend) respectively. $m$, $l$ and $n$ denote the first few bound states in the three well widths. Note that the resonances occur at slightly smaller energies than that obtained from equation~\eqref{eq22}. Broken vertical line denotes $\kappa=\sqrt{V_0}\sim 6.32$.}
\label{fig3}
\end{figure}

Instead, if one takes another 4BP which has the 3 wells of different 
widths (i.e. an asymmetric 4BP), the resonances must be distinctly resolved 
at all energies. This was also observed by Rao. et al. for a 3BP.\cite{sid3} We 
illustrate this feature in Fig.~\ref{fig3}(a), for an asymmetric four barrier of 
constant barrier width $\delta =0.5 $, height $V_{0} = 40$, but different 
well widths $\tau_{1} =5,\ \tau_{2} =3,\ \tau_{3} =2$. 

\begin{figure*}[p]
\centering
\includegraphics[scale=0.5]{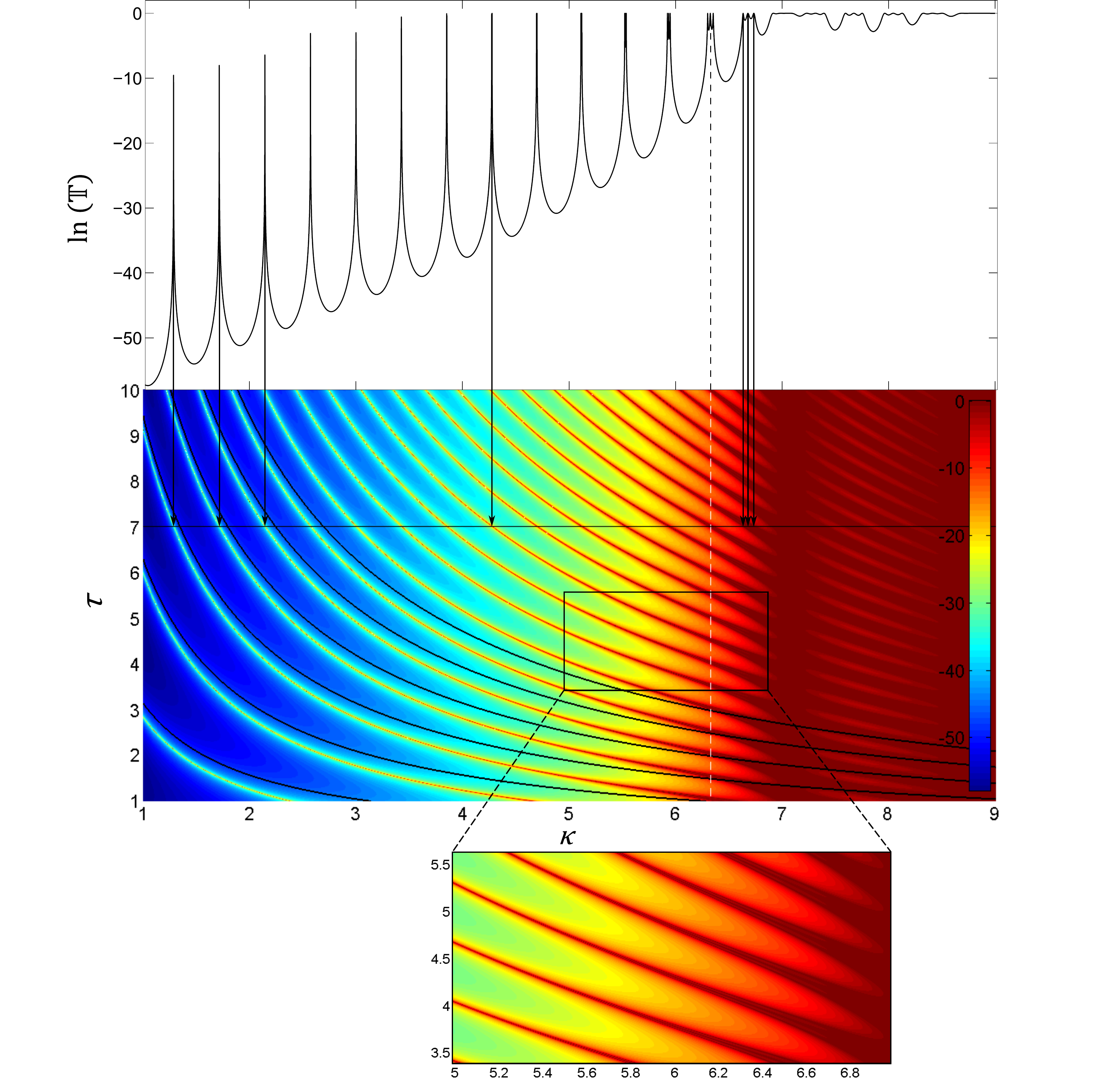}
\caption{(top) $\ln(\mathbb{T})$ vs. $\kappa$ for 4BP of $V_0 = 40$, $\delta= 1$ and $\tau= 7$. (middle) Color plot
of $\ln(\mathbb{T}(\kappa, \tau))$ on the $\kappa$--$\tau$ plane. Color legend gives the value of $\ln(\mathbb{T})$. 
Broken vertical line (white) denotes $\kappa= \sqrt{V_0}$ while solid horizontal line (black) denotes
$\tau = 7$. Note that $\mathbb{T}$ along this line ($= \mathbb{T}(\kappa, 7)$) pertains to the barrier geometry of
the top figure. Hence there is a one to one correspondence between the resonant
peaks (some are mapped by vertical arrows). Black solid curves show rectangular
hyperbolas $\kappa\tau = n\pi$ for $n = 1,2,3,\ldots$. These curves embrace the resonant tracks for
$\kappa \ll V_0$. (bottom) Magnified view of the boxed region of the middle figure, exhibiting
the trifurcation of the resonant tracks.}
\label{fig4}
\end{figure*}

\begin{figure*}[p]
\centering
\vspace*{100pt}
\includegraphics[scale=0.5]{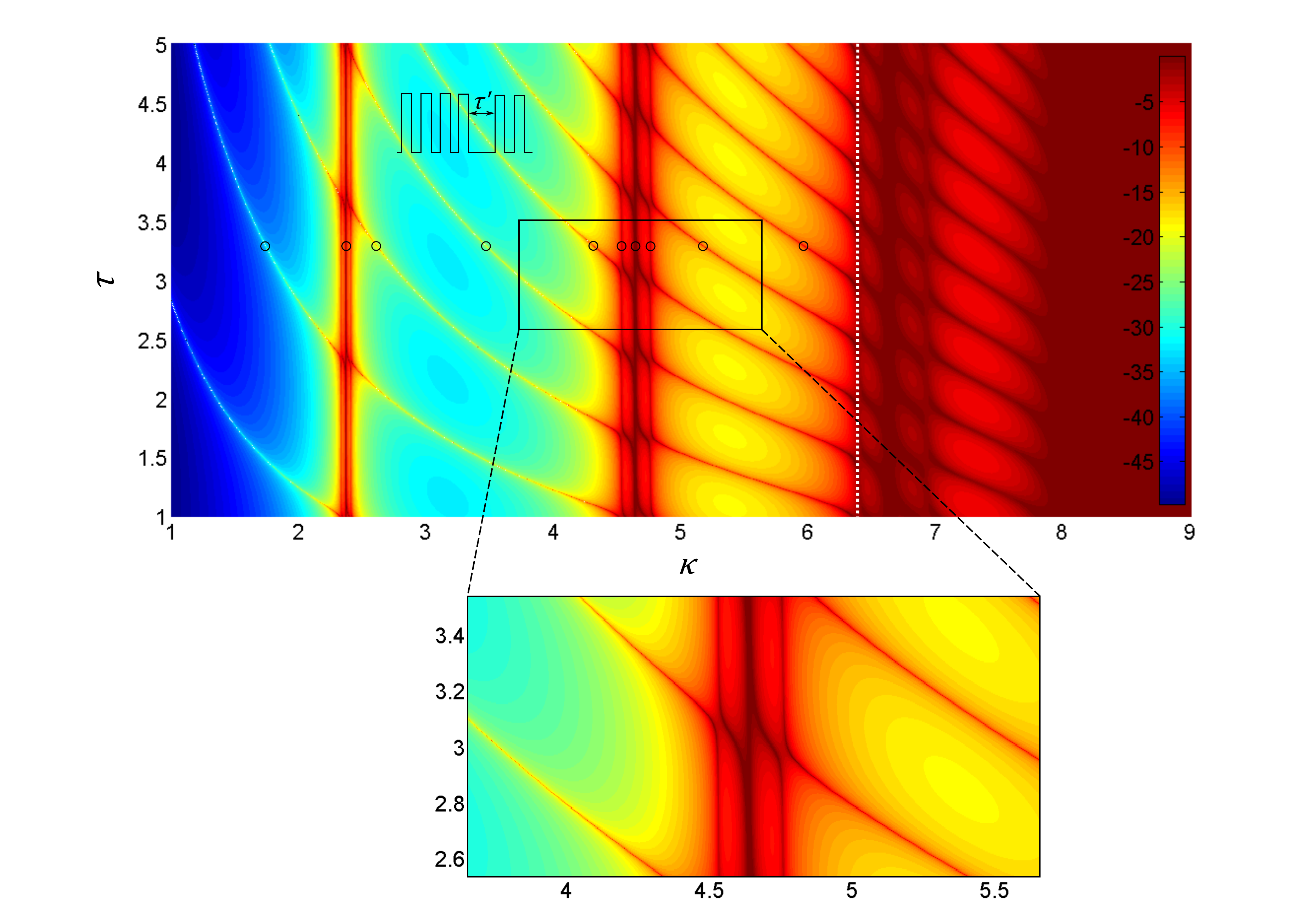}
\caption{
Color plot of $\ln(\mathbb{T}(\kappa, \tau'))$ for an asymmetric 6BP. Color bar gives the probability
scale. All barriers are of same height $V_0 = 40$, and width $\delta = 1$. All wells,
except the fourth well have width $\tau=1$. The fourth well has a width  $\tau'$ which is
varied as a parameter along the vertical axis from 1 to 5. $\kappa$ is taken along horizontal
axis. The barrier is pictured schematically in the figure. Note that, due to
 alias effect (discussed later), it would not matter if any other well width
was chosen as $\tau'$ (instead of the fourth well). The picture would essentially remain
intact for the energy ranges considered here. Also the bottom of the picture $(\tau' = 1)$
pertains to the uniform 6BP.
}
\label{fig5}
\vspace*{100pt}
\end{figure*}

The barrier is also pictured as an inset.
Due to the inverse relation between $\kappa $ and $\tau $ in equation~\eqref{eq22}, 
the resonant peaks are ordered in a specific manner, i.e. the well 
with the maximum width contributes to the \textit{initial} resonant peaks.

The resonant peaks above $\kappa =\sqrt V_{0} $ have broadened. This is 
expected from the classical tunneling characteristics\footnote{By classical tunneling we refer to 
$\mathbb{T}=\begin{cases} 
 0 &\kappa <\sqrt V_{0} \\ 
 1 &\kappa >\sqrt V_{0}. 
 \end{cases} $} 
(which sets in as 
the large energy limit of the quantum mechanical transmission 
coefficient). In Fig.~\ref{fig3}(b) we plot $\ln (\mathbb{T})$𝕋 vs. 𝜅$\kappa$
for the asymmetric 4BP (same as Fig.~\ref{fig3}(a)) 
along with that of \textit{uniform} 
4BPs of same  barrier height 𝑉$V_0= 40$, barrier width 𝛿$\delta = 1$\footnote{We choose $\delta =1$ (instead of 0.5) for the colored plots, to 
bring out the correspondence better. If one takes $\delta =0.5$ the curves 
would overlap. A higher value of delta lowers the value of $\mathbb{T}$ and has \textit{no} 
effect on the position of the peaks. Hence the correspondence between the 
resonant peaks of the asymmetric MBP and the uniform MBPs is still 
preserved.}
and well widths $\tau =5$ (red), $\tau =3$ (green) and $\tau =1$ (blue). The most striking 
feature is that there is a \textit{one to one} correspondence 
between the resonant peaks of the uniform barrier plots and those of the asymmetric barrier. This is a 
reaffirmation of our previous remark---every well contributes its resonant 
energies \textit{independently}, in accordance with equation~\eqref{eq22}. In Fig.~\ref{fig3}(b) the resonances 
are labeled by quantum numbers $m$, $n$ and $l$ for the three different wells. Note, 
in Fig.~\ref{fig2}(b) the wave function corresponded to an energy of 27.217, at which 
the first resonant peak of the \textit{fourth} band arises. The wave function for this 
case becomes nearly sinusoidal at the site of the wells and has 3 well 
defined nodes in \textit{each} well, which is an attribute of the fourth bound state wave 
function of a particle in a box of width $\tau $.

A natural way to extend equation~\eqref{eq23} for a generic MBP of varying well 
widths $\tau_{j} $, $j=1,2,\ldots m-1$ (but same barrier height 
$V_{0})$ is found:\footnotemark[4]
\begin{equation}\label{eq24}
\alpha =\sum\limits_{r=1}^R \left[\frac{\tau_{a_{r}} \sqrt V_{0}} {\pi 
} \right],\ a_{r} \in \left\{1,2,\ldots m-1\right\}, \  
 R\le m-1. 
\end{equation}
Well widths that are \textit{repeated} must be counted only once, since they give the same 
resonant set. $a_{r} $ takes care of the distinct well widths, $R$ being 
the total number of such widths.\footnote{For $n\tau = l\tau'$ the peaks would overlap and this must be 
accounted for in equation~\eqref{eq24}. But resonances do not \textit{strictly} follow equation~\eqref{eq22}. Thus the formula works and this correction is not needed.}

For the illustrative asymmetric 4BP 
we have $\alpha =\sum\limits_{j=1}^3 \left[\frac{\tau_{j} \sqrt {40}} {\pi 
} \right] =23$. Figure~\ref{fig3}(a) gives 20 peaks (below $V_{0})$ and 3 diffuse peaks 
around $V_{0} $ (vertical broken line)---a total of 23!

 Clearly, the well widths of a MBP play a very special role in positioning 
the resonant peaks. This is explored further in Fig.~\ref{fig4}, which takes $\tau $ 
(the well width of a \textit{uniform} MBP) as an independent parameter and plots 
$\ln \left(\mathbb{T}(\kappa ,\tau)  \right)$ on the $\kappa 
-\tau $ plane for $m=4$, $V_{0} =40$, and $\delta =1$. The color scale 
gives the value of $\ln (\mathbb{T})$. The horizontal axis is 
$\kappa $ and $\tau $ ranges from 1 to 10 on the vertical axis. The 
broken vertical line (white) is $\kappa =\sqrt V_{0} $. Resonant peaks 
project out as red dots, defining \textit{distinct tracks} on the $\kappa-\tau $ plane. These 
tracks begin as isolated curves at low energies, gradually branch into 3 
tracks, at middle energies, entering the region $\kappa >\sqrt V_{0} $. 
(Fig.~\ref{fig4} inset gives a magnified view of these branches) In general we expect 
$m-1$ tracks for a \textit{uniform} MBP. The branching of the \textit{resonant tracks} is a consequence of the resolution of the $m-1$ peaks in each band with increasing energy observed 
earlier. 

 Consider the horizontal broken line (black) at $\tau =7$. This corresponds 
to a uniform 4BP of well width 7 and other parameters are same as above. 
(The transmission characteristics of this 4BP is provided in the top of 
Fig.~\ref{fig4}). Note that the intersections of the horizontal line with the resonant 
tracks (for $\kappa <\sqrt V_{0})$ is consistent with the resonant peaks of 
the top figure (mapped by means of vertical arrows). 
 From a mathematical standpoint the red regions display a smooth 
continuation from the \textit{discrete} bound states of the infinite square well (in the 
left) to the \textit{continuum} states of the free particle (right), where the resonant tracks 
coalesce into a continuous `band' for $\kappa >\sqrt V_{0} $. Superposed on 
Fig.~\ref{fig4} are rectangular hyperbolas (black continuous curves) defined in 
equation~\eqref{eq22} for $n=1,2,3,\ldots$ which approximate the resonant 
tracks (for small $\kappa)$ and later deviate as $\kappa$ becomes 
comparable to $\sqrt V_{0} $. In fact, this picture sets the regime of 
energies for which asymptotic analysis using the results of the infinite 
square well problem are valid. The number of intersections of a horizontal 
line (at a given $\tau)$ with the resonant tracks below $\kappa =\sqrt 
V_{0} $ gives the number of 
resonant bands ($=\beta)$. The direct 
proportionality between $\beta $ and $\tau $ (equation~\eqref{eq23}) is captured 
in Fig.~\ref{fig4} i.e. horizontal lines at smaller $\tau $ have lesser number of 
intersections with the resonant tracks compared to those at larger $\tau $. 
Analysis of the tunneling characteristics of a MBP on the $\kappa$--$\tau $ 
plane is very insightful. It projects the role of the well width $\tau $, in 
a natural way.

\begin{figure}[h!]
\centering
\includegraphics[scale=0.3]{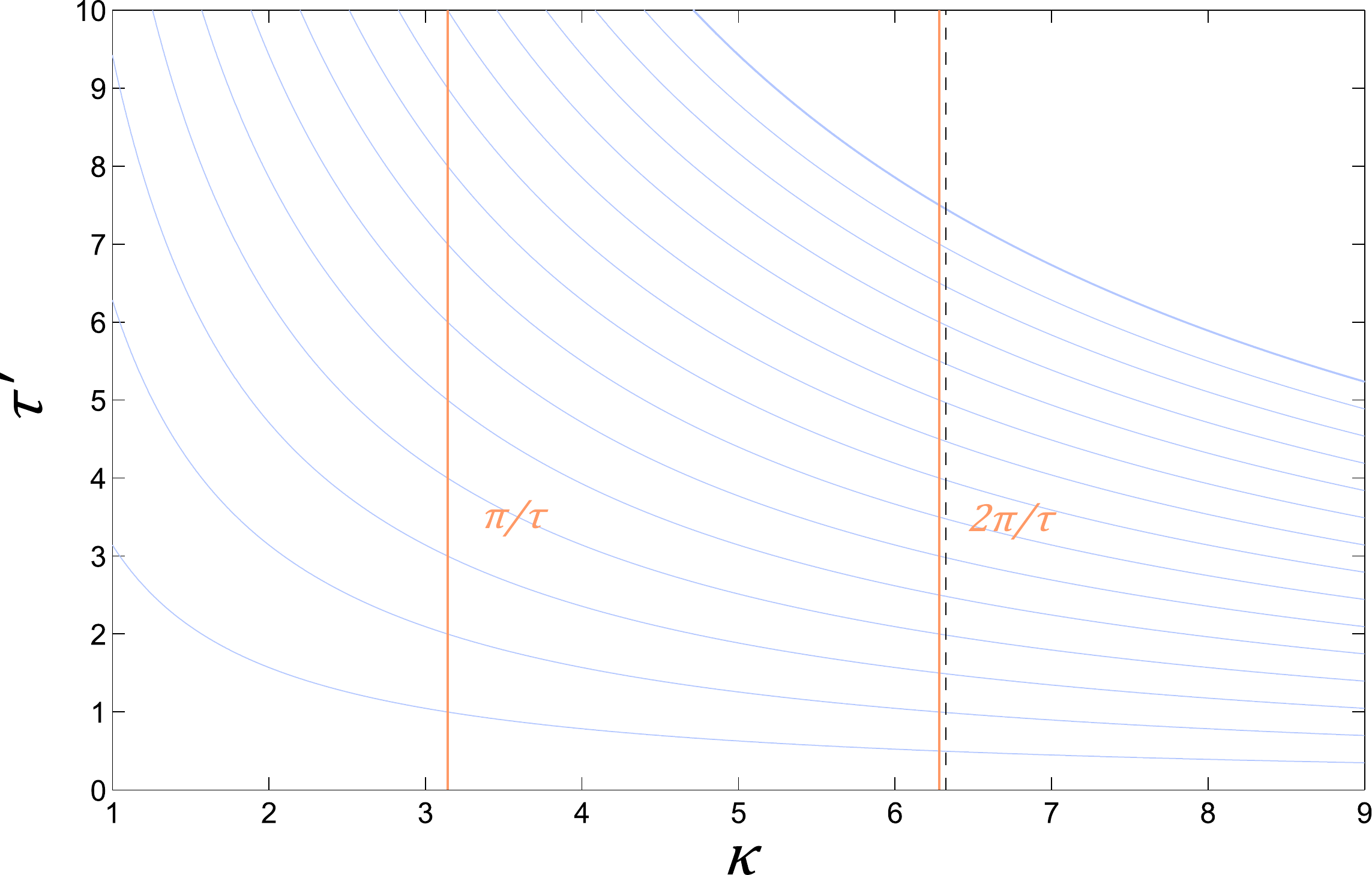}
\caption{Vertical red lines are the first two resonances of the well width $\tau$ 
(that can be accommodated below $V_0$) while the blue curves are the resonant 
tracks generated by the resonances of the well width $\tau'$ by the equation(s)  
$\kappa\tau'= l\pi,\ l=1,2,3,\ldots,15$.}
\label{fig6}
\end{figure}

Now we consider an asymmetric 6BP of barrier specifications $V_{0} =40$, 
$\delta =1$. For the plot of Fig.~\ref{fig5} we have taken the well widths of all the 
wells (except the fourth one) $\tau =1$. The fourth well has a width  
$\tau '$ which is varied as a parameter on the vertical axis from 1 to 
5. As in the previous figure, the color denotes the logarithm of the 
transmission coefficient and $\kappa $ is taken along the horizontal axis. 
Note that there is nothing special about the fourth well and any other well 
width could be chosen as $\tau '$ (without altering the picture appreciably). 
This freedom is attributed to the \textit{alias effect} 
discussed later. The barrier is sketched schematically in the figure. 
Clearly, the dynamics of the resonant peaks in this case is extremely 
non-trivial and a complex structure emerges even with one well perturbation.\cite{sid9} Although, the exact details of the figure are quite perplexing, the 
basic frame of the resonant tracks can be reasoned in a simple manner using 
equation~\eqref{eq22}. The well width $\tau $ contributes resonances at $\kappa 
=\frac{n\pi} {\tau} $. Below the barrier top ($\kappa <V_{0})$ only about 2 
bands can be accommodated for $\tau =1$. (i.e. $\beta =2)$ Thus these 
resonant peaks define vertical tracks in the figure. This also implies that, 
irrespective of the value of $\tau '$, the resonances due to the width $\tau 
$ will always show up. The curved resonant tracks that percolate through 
these vertical lines are due to the well width $\tau '$ obtained from: 
$\kappa \tau' =l\pi,\  l=1,2,3,\ldots $ If we proceed just 
with the asymptotic formula (equation~\eqref{eq22}), we would contrive a picture 
that looks like Fig.~\ref{fig6}. Clearly, Fig.~\ref{fig6} does not describe the 
actual situation. This is primarily because, equation~\eqref{eq22} is valid at 
energies well below the barrier top. Note that the 
vertical resonant tracks (red) do not show splitting (in Fig.~\ref{fig6}) which 
actually occurs (in Fig.~\ref{fig5}) due to coupling effects, discussed earlier. 

Apart from details at the intersection points (which are vital) Fig.~\ref{fig6} 
portrays the network of resonant tracks of Fig.~\ref{fig5} fairly accurately. In Fig.~\ref{fig6} 
the intersections correspond to points where $n\tau =l\tau',\ n, l\ne 
0$. If there was a true intersection then it would lead to an overlap of 
resonant spikes.\footnotemark[7] This doesn't occur in the actual picture (Fig.~\ref{fig5}) 
since the asymptotic formulas become inaccurate at intermediate energies 
(Ref. Fig.~\ref{fig5} inset). In any case equation~\eqref{eq24} holds good. For instance 
consider the well width $\tau^{'} =3.35$ at which the number of resonant 
spikes must be 10. This is validated in Fig.~\ref{fig5} with  black circles 
encircling  the resonant peaks. Also the resonant peaks are more densely 
distributed around regions where the tracks due to both the well widths come 
very close and their population drops in the interstices. The bottom of 
Fig.~\ref{fig5} (where $\tau' =1)$ corresponds to the symmetric or uniform 6 
barrier problem. As the perturbation is gradually turned on, distortions set 
in and there arises a `cross talk' between the hyperbolic tracks of ($\tau 
')$ via the stationary resonances of ($\tau)$ which are vertical. When 
observed closely it is found that the hyperbolic tracks smoothly \textit{deform into 
each other} at the site of the vertical resonant tracks. This is a remarkable 
feature of the asymmetric multi-barrier problem.

\subsection{Permutation invariance and alias effect} \label{sec3.1}
The transmission coefficient plots become particularly interesting for a 
specific class of MBPs. We have seen that the well widths have a strong 
bearing on the position of the resonant peaks. The role of well widths is 
expounded further in this sub section. Consider an asymmetric multi-barrier 
system of barrier height $V_{0} $, barrier width $\delta $, and well widths 
$\tau_{1} $, $\tau_{2} $,\ldots$\tau_{m-1} $ (starting from left). If the 
ordering of the wells is `ignored'-then one can construct more asymmetric 
multi-barrier potentials (using the parameters of the above prototype) by 
permuting the position of the wells. If all the well widths are distinct, 
there are $\left(m-1 \right)!$ possible MBPs. We refer to these barriers as 
\textit{permutation-equivalent} MBPs. These MBPs possess the same set of well widths. Thus the transmission 
characteristics must always have the \textit{same number} of resonant peaks in accordance with 
equation~\eqref{eq24}. Moreover, equation~\eqref{eq22} guarentees that the resonances 
contributed by each $\tau_{j} $ would occur at the \textit{same} \textit{location}. Even if the exact 
position of the resonances are influenced by the barrier height and width, 
that influence would be the same for all the permutation-equivalent MBPs. 
Thus the actual position of the resonances must indeed be the \textit{same} for these 
multi-barrier systems.

\begin{figure}
\centering
\subfigure{(a)}{\includegraphics[scale=0.25]{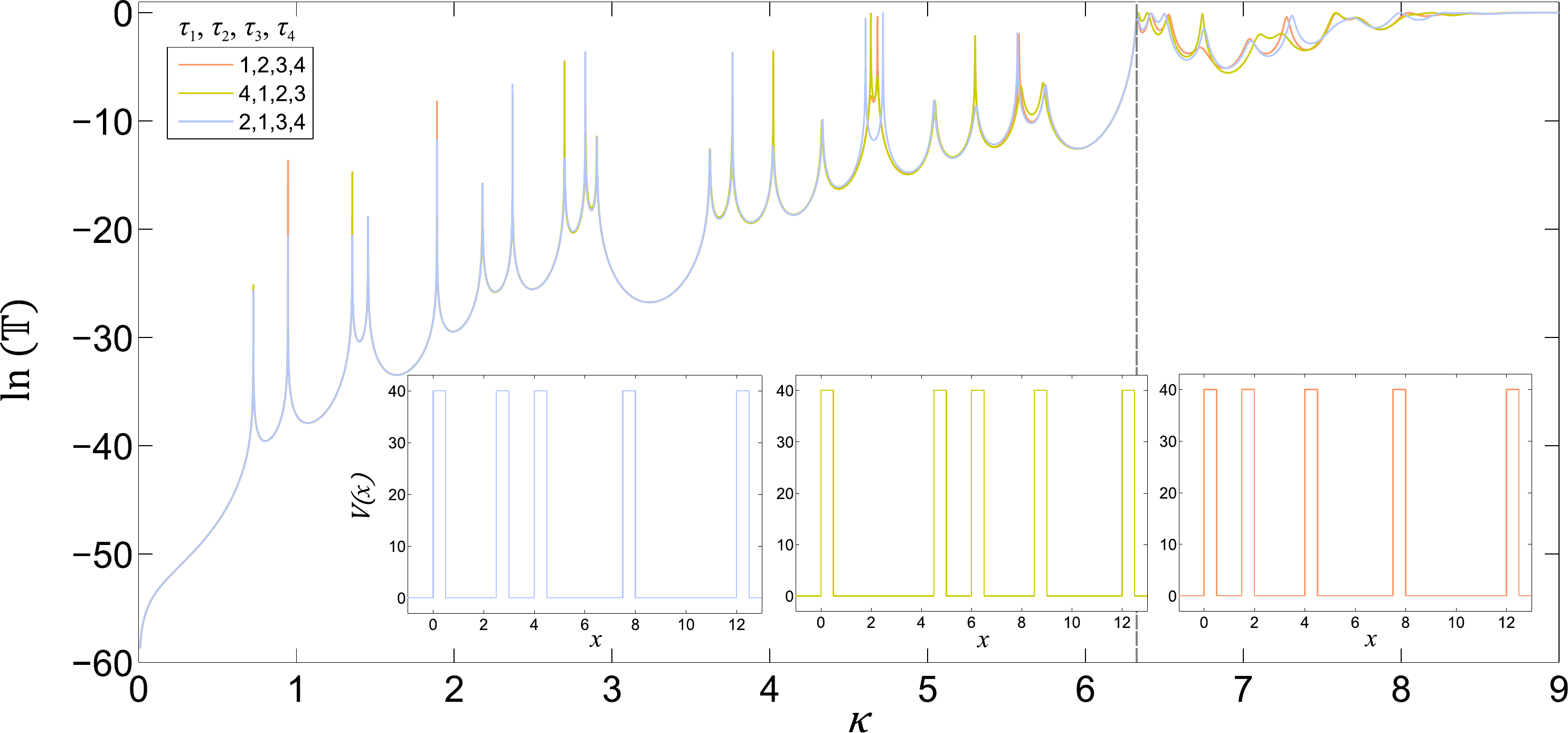}}
\subfigure{(b)}{\includegraphics[scale=0.25]{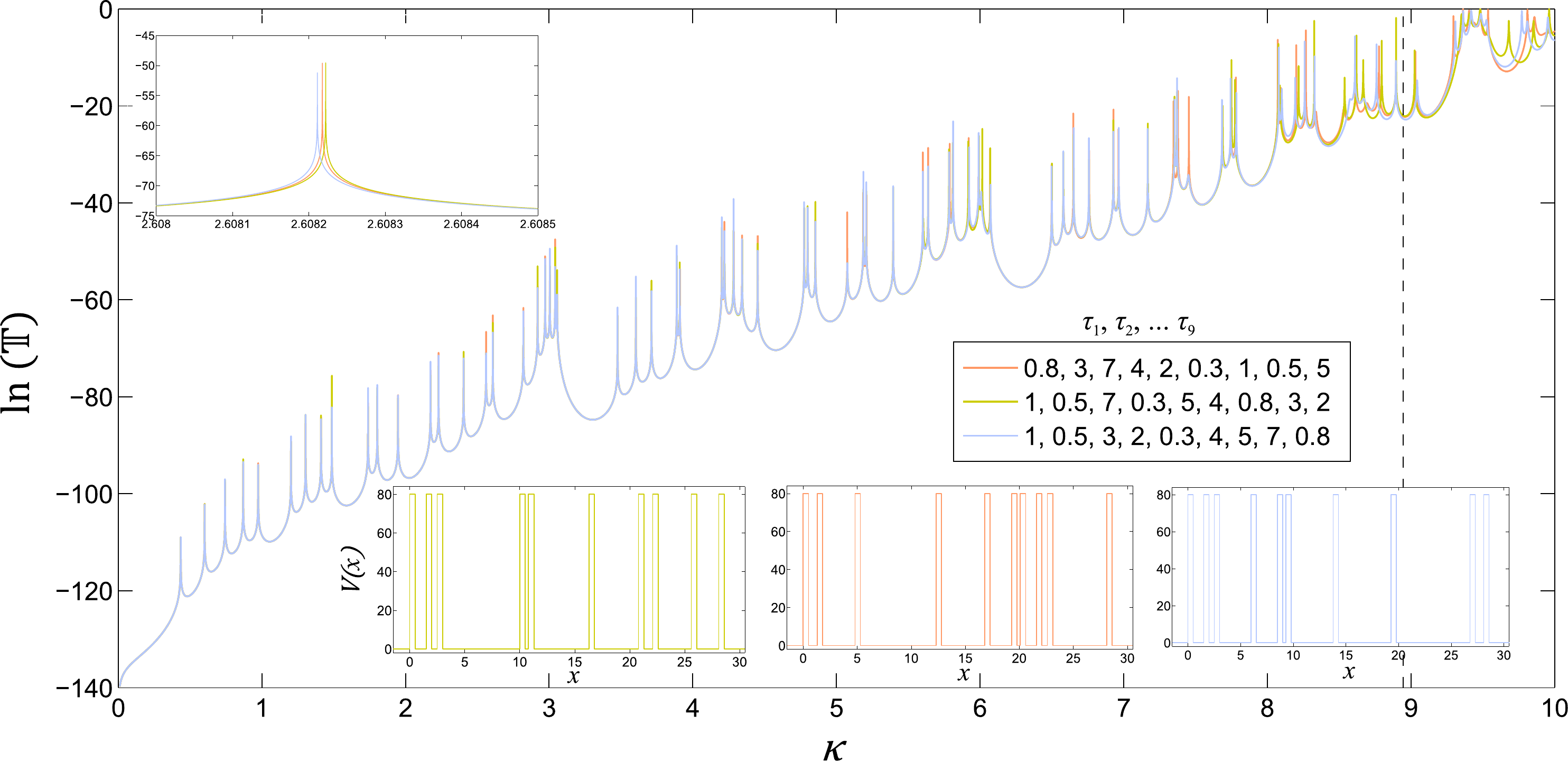}}
\caption{(a) $\ln(\mathbb{T})$ vs. $\kappa$ for three asymmetric 5BPs with $V_0=40,\ \delta=0.5,\ \tau_1, \tau_2, \tau_3, \tau_4 \in \{1,2,3,4\}$. The legend gives the well widths for each plot in the specified order. Notice the aliasing of the three transmission curves for $\kappa <\sqrt{V_0}$. (b)  $\ln(\mathbb{T})$ vs. $\kappa$ for three asymmetric 10BPs with same barrier parameters as Fig.~\ref{fig7}(a). Well widths $\tau_1, \tau_2, \ldots, \tau_9 \in \{1,0.5,3,0.3,2,5,4,0.8,7\}$. Corresponding  $V(x)$s are shown as insets. (inset) magnified view of 17th resonant peak.} 
\label{fig7}
\end{figure}

Here we have implicitly assumed that the actual ordering of the wells 
doesn't influence the position of the resonant peaks. We will see that this 
assumption will get challenged later. Thus it seems like, the transmission 
coefficients of these barriers would have some sort of similarity (at least 
in the region where equation~\eqref{eq22} is valid). We illustrate this 
pictorially in Fig.~\ref{fig7}(a) with an asymmetric 5BP of barrier height $V_{0} =40$, 
width $\delta =0.5$ with different well widths (labeled from left to right --
$\tau_{1}, \tau_{2}, \tau_{3}, \tau_{4})$. We have chosen $1, 
2, 3, 4$ as the widths and graph three permutations of these widths 
pertaining to three different asymmetric 5BPs. $V(x)$ for the corresponding 
curves are depicted as insets. 
 
Quite consistent with our expectation, the curves nearly overlap! At 
energies below the barrier top, the curves (pertaining to the different 
spatial permutations of the well widths) get superposed and the differences 
between them surfaces only at higher energies. We refer to this phenomenon 
as the \textit{Permutation-Invariant Alias Effect} (or \textit{Alias-Effect}). 
Note that for a MBP there would be $\frac{1} {2} (m-1)!$ aliased 
solutions.\footnote{Although we have $\left(m-1 \right)!$ distinct \textit{permutation equivalent} MBPs, some of them are 
not aliased solutions. For instance in a 4BP with three distinct well widths 
$\tau_{1} $, $ \tau_{2} $ and $\tau_{3} $, consider the permutations $\tau 
_{3}$--$\tau_{1}$--$\tau_{2} $ and $\tau_{2}$--$\tau_{1}$--$\tau_{3} $. They 
would result in the same transmission coefficient, since a particle incident 
from left on the first barrier is equivalent to a particle incident from  
right on the other. And the transmission coefficient of the barrier must not 
depend on the direction from which the particle approaches. In fact this 
feature is inherently embedded in the structure of the transfer matrices and 
can be rigorously proven.}

Aliasing of the transmission coefficients is not lost even if the barrier number is increased significantly. 
In Fig.~\ref{fig7}(b) we plot the case of a 10BP. The barrier 
parameters are same as that of Fig.~\ref{fig7}(a). 
and the well widths are listed in the legend. Note that the aliasing is not 
perfect at the site of the resonant peaks. In fact the separation occurs on 
a very narrow range of energy. (Fig.~\ref{fig7}(b) inset). This is attributed to the 
coupling, which might depend on the ordering of the wells. We conclude by 
rephrasing the Aliasing condition.

\textit{The transmission characteristics of an asymmetric MBP is `invariant' to the spatial permutation of the} $m-1$ \textit{wells so long as the heights and the widths of the} $m$ \textit{barriers are kept same}.

\section{Conclusion} \label{sec4}

In the discussion of asymmetry we focused primarily on the well widths. This 
doesn't imply that the barrier widths or heights wouldn't play a big role. If 
the barrier height is increased, the threshold gets shifted further, while
the overall behavior is not affected appreciably. So far as barrier widths 
are concerned---an increase simply leads to a lowering of the over all 
probability, however the resonant peaks are not significantly affected. The 
width of the barriers have a small bearing on the extent to which the spikes 
(in each band) are resolved. At any rate, it is the well widths that 
completely dictate the transmission characteristics of a MBP. We have thus 
applied our formulation  to study asymmetry in a multi-barrier structure. As 
noted earlier, these are special examples of piecewise constant potential 
barriers. And there are several other applications to which the methods 
developed in this paper can be used. We reserve a discussion of some of 
these for future papers. At this point we emphasize the importance of 
equation~\eqref{eq12} which was a crucial ingredient in the solution. 
 
Some of the topics that have not been considered are tunneling time and 
tunneling length. These are interesting parameters to look at for a MBP. 
Time evolution of the wave function is another aspect that requires further 
insight. Certainly the analysis of these problems rests directly on the 
discussion provided in this paper. Also, long sequence matrix products of 
the form presented here, call for optimal computational algorithms that 
reduce code and time complexities.

\begin{acknowledgments}
The author extends his acknowledgement to Dr. Pankaj Agarawal (Institute of 
Physics, Bhubaneswar), Dr. Hemalatha Thiagarajan (NIT, Trichy), Dr. S.~D. 
Mahanti (Michigan St. University), Dr. T.~N. Janakiraman (NIT, Trichy) for 
the many ways in which they have contributed to the completion of this 
paper. Sambhav R. Jain (Design Engineer, Texas Instruments) has prepared the 
excellent graphics for this paper.  Special thanks are due to Prof. D.~J. Griffiths 
for encouraging publication of this work and to the anonymous reviewers for their valuable suggestions.
\end{acknowledgments}
\vfill\eject

\bibliography{abhijit}
\end{document}